\documentclass[prl,nofootinbib,twocolumn]{revtex4}

\usepackage{amssymb}
\usepackage{amsmath}
\usepackage[dvips]{graphicx}
\usepackage{longtable}
\usepackage{verbatim}
\usepackage{amsfonts}

\newcommand{\be}{\begin{equation}}
\newcommand{\ee}{\end{equation}}
\newcommand{\bea}{\begin{eqnarray}}
\newcommand{\eea}{\end{eqnarray}}
\newcommand{\beq}{\begin{equation}}
\newcommand{\eeq}{\end{equation}}
\def\beqa{\begin{eqnarray}}
  \def\eeqa{\end{eqnarray}}
\newcommand{\bv}{\left(\begin{array}{c}}
\newcommand{\ev}{\end{array}\right)}
\newcommand{\no}{\nonumber}
\def\lsim{\mathrel{\rlap{\lower4pt\hbox{\hskip1pt$\sim$}}
    \raise1pt\hbox{$<$}}}         
\def\gsim{\mathrel{\rlap{\lower4pt\hbox{\hskip1pt$\sim$}}
    \raise1pt\hbox{$>$}}}         

\begin{document}

\begin{flushright}
DO-TH 12/09
\end{flushright}

\vspace*{-30mm}

\title{\boldmath Supersymmetric $\Delta A_{CP}$}

\author{Gudrun Hiller}\email{ghiller@physik.uni-dortmund.de}
\affiliation{Institut f\"ur Physik, Technische Universit\"at Dortmund, D-44221 Dortmund, Germany}

\author{Yonit Hochberg}\email{yonit.hochberg@weizmann.ac.il}
\affiliation{Department of Particle Physics and Astrophysics,
  Weizmann Institute of Science, Rehovot 76100, Israel}

\author{Yosef Nir}\email{yosef.nir@weizmann.ac.il}
\affiliation{Department of Particle Physics and Astrophysics,
  Weizmann Institute of Science, Rehovot 76100, Israel}

\vspace*{1cm}

\begin{abstract}
  There is experimental evidence for a direct CP asymmetry in singly
  Cabibbo suppressed $D$ decays, $\Delta A_{CP}\sim0.006$. Naive
  expectations are that the Standard Model contribution to $\Delta
  A_{CP}$ is an order of magnitude smaller.  We explore the
  possibility that a major part of the asymmetry comes from
  supersymmetric contributions. The leading candidates are models
  where the flavor structure of the trilinear scalar couplings is
  related to the structure of the Yukawa couplings via approximate
  flavor symmetries, particularly $U(1)$, $[U(1)]^2$ and $U(2)$.  The
  recent hints for a lightest neutral Higgs boson with mass around 125
  GeV support the requisite order one trilinear terms.  The typical
  value of the supersymmetric contribution to the asymmetry is $\Delta
  A_{CP}^{\rm SUSY}\sim 0.001$, but it could be accidentally enhanced
  by order one coefficients.
\end{abstract}

\maketitle

\section{Introduction}
The world average for the direct CP asymmetry in singly Cabibbo
suppressed $D$ decays, based on measurements by E687
\cite{Frabetti:1994kv}, CLEO \cite{Bartelt:1995vr,Csorna:2001ww}, E791
\cite{Aitala:1997ff}, FOCUS \cite{Link:2000aw}, BaBar
\cite{Aubert:2007if}, Belle \cite{:2008rx}, CDF
\cite{Aaltonen:2011se,CDFLaThuile2012} and LHCb
\cite{Aaij:2011in}, is now $4.3\sigma$ away from zero \cite{HFAGMarch2012}:
\beqa\label{eq:acp}
\Delta A_{CP}&\equiv&
A_{CP}(K^+K^-)-A_{CP}(\pi^+\pi^-)\no\\
&=&-0.00656\pm0.00154.
\eeqa
Here,
\beq
A_{CP}(f)=\frac{\Gamma(D^0\to f)-\Gamma(\overline{D}^0\to f)}
{\Gamma(D^0\to f)+\Gamma(\overline{D}^0\to f)}.
\eeq
In $\Delta A_{CP}$, that is the difference between asymmetries,
effects of indirect CP violation largely cancel out
\cite{Grossman:2006jg}. Thus, $\Delta A_{CP}$ is a
manifestation of CP violation in decay.

The Standard Model (SM) contribution to the individual asymmetries is
suppressed by a CKM factor of order $2{\cal
  I}m\left(\frac{V_{ub}V_{cb}^*}{V_{us}V_{cs}^*}\right)
\approx1.2\times10^{-3}$, and by a loop factor of order
$\alpha_s(m_c)/\pi\sim0.1$.  While one cannot exclude an enhancement
factor of order 30 from hadronic physics
\cite{Golden:1989qx,Brod:2011re,Pirtskhalava:2011va,Cheng:2012wr,
Bhattacharya:2012ah,Li:2012cf,Franco:2012ck,Feldmann:2012js,Brod:2012ud}, in
which case (\ref{eq:acp}) will be accounted for by SM physics, it is
interesting to explore the possibility that new physics contributes a
major part of $\Delta A_{CP}$.

The size of new physics contributions to $\Delta A_{CP}$ is often
constrained by other flavor-related observables, such as
$D^0-\overline{D}^0$ mixing or $\epsilon^\prime/\epsilon$
\cite{arXiv:1111.4987}. Supersymmetric models, via their contribution
to the chromomagnetic operator, can generate large enough asymmetry in
$D$ decays without conflicting with these observables
\cite{Grossman:2006jg,Giudice:2012qq,Altmannshofer:2012ur}. In this work, we investigate
whether this scenario is likely to be realized in supersymmetric
models with viable and natural flavor structure.

\section{The supersymmetric parameters}
The $6\times6$ mass-squared matrix for the up- and down-type squarks
can be decomposed into $3\times3$ blocks, $q=u,d$,
\beq \tilde M^{2q}=\left(\begin{array} {cc}
    \tilde M^{2q}_{LL} & \tilde M^{2q}_{LR} \\
    \tilde M^{2q}_{RL} & \tilde M^{2q}_{RR} \end{array}\right) +
D,F\mbox{-terms}, \eeq
where $L$ and $R$ denote SU(2) doublets and singlets, respectively.
We denote the average squark mass by $\tilde m$. Then, it is
convenient to parameterize the supersymmetric contributions to flavor
changing processes in terms of dimensionless parameters,
\beq
(\delta^q_{MN})_{ij}=\frac{(\tilde M^{2q}_{MN})_{ij}}{\tilde m^2},
\eeq
where $M,N=L,R$. When, to a good approximation, only two squark generations are
involved, one can express these parameters in terms of the
supersymmetric mixing angles, $(K^q_M)_{ij}$, and the mass-squared
splittings between squarks, $\Delta \tilde m^2_{ij}$:
\beq (\delta^q_{MN})_{ij}=\frac{\Delta\tilde
  m^2_{q_{Mi}q_{Nj}}}{\tilde m^2}(K^q_M)_{ij}(K^q_N)_{jj}.  \eeq

The parameters that are most relevant to $\Delta A_{CP}$ are
$\delta_{LL}\equiv(\delta^u_{LL})_{12}$ and
$\delta_{LR}\equiv(\delta^u_{LR})_{12}$, which generate the
chromomagnetic operator with Wilson coefficient given by
\beq
C_{8g}=F(x)\delta_{LL}+G(x)\frac{m_{\tilde g}}{m_c}\delta_{LR},
\eeq
where $x=(m_{\tilde g}^2/\tilde m^2)$, and the functions $F$ and $G$
can be found, for example, in Ref. \cite{Grossman:2006jg}. Given that
$G(x)$ is larger than $F(x)$ by a factor of a few, and the enhancement
factor of $m_{\tilde g}/m_c$, the dominant contribution in the models
that we consider comes from $\delta_{LR}$.  It can be estimated as
follows \cite{Giudice:2012qq}:
\beq
\Delta A_{CP}^{\rm SUSY} \sim 0.006\ \frac{{\cal I}m(\delta_{LR})}{0.001}\
\frac{1\ {\rm TeV}}{\tilde m}.
\eeq
In the following sections, we investigate whether
\beq\label{eq:desired}
{\cal I}m(\delta_{LR})\sim0.001
\eeq
can plausibly arise in supersymmetric flavor models.

\section{Supersymmetric flavor models}
If the soft supersymmetry breaking terms had a generic flavor
structure (``anarchy"), the supersymmetric contributions to flavor
changing neutral current processes would exceed experimental
constraints by orders of magnitude. Thus, these terms must have a
special structure.

The most extreme solution to this ``supersymmetric flavor puzzle" is a
constrained version of minimal flavor violation (MFV): at the supersymmetry
breaking mediation scale, squark masses are universal and the trilinear scalar 
couplings (the $A$ terms) are proportional to the corresponding Yukawa matrices. 
Such a situation arises naturally in various mediation schemes, most notably gauge- 
and anomaly-mediated supersymmetry breaking.  The renormalization group
evolution does generate flavor changing effects in the soft breaking terms, but
MFV implies a very strong flavor suppression of these effects,
\beq
\delta_{LR} \propto \frac{m_c}{\tilde m}(V_{us} V_{cs}^* y_s^2 +
V_{ub} V_{cb}^* y_b^2 ) \lesssim {\cal{O}}(10^{-7}).
\eeq
(For the exact expression in anomaly mediation, see
\cite{Allanach:2009ne}.)

It is possible, however, that the supersymmetric flavor structure is
related to that of the Standard Model, but is not MFV. This is the
case in models where an approximate flavor symmetry dictates the
structure of all flavor changing couplings. In what follows, we
examine several such symmetries -- $U(1)$, $[U(1)]^2$, $U(2)$ and
$[U(2)]^3$ -- with regard to their implications for ${\cal
  I}m(\delta_{LR})$.

\subsection{Abelian Symmetries}
The Froggatt-Nielsen (FN) framework \cite{Froggatt:1978nt} postulates
an approximate $U(1)$ symmetry, broken by a spurion with value $\ll1$.
Assigning different charges to the different quark generations results
in parameterically suppressed quark mass ratios and mixing angles. In
supersymmetric FN models \cite{Nir:1993mx,Leurer:1993gy}, the squark
spectrum is anarchical, up to some level of degeneracy between the
first two generations from renormalization group evolution (RGE)
effects, but the mixing angles are small.

With a single $U(1)$ symmetry, the parametric suppression of squark
flavor parameters is related to quark flavor parameters, independent
of details of the model such as the size of the spurion and the charge
assignments. In particular, the following relations hold for the
entries that are relevant to $c\to u$ transitions \cite{Nir:2002ah}:
\beqa
(\delta^u_{LL})_{12}&\sim&\frac{|V_{us}|}{r_3},\label{eq:LL}\\
(\delta^u_{RR})_{12}&\sim&\frac{m_u}{r_3 m_c|V_{us}|},\label{eq:RR}\\
(\delta^u_{LR})_{12}&\sim&\frac{\tilde a}{\tilde m}\frac{m_c|V_{us}|}{\tilde m},
\label{eq:LR}
\eeqa
where $\tilde a$ is the typical scale of the $A$-terms. The $1/r_3$
factor, defined in Ref. \cite{Hiller:2008sv}, represents the
gluino-related RGE effect which generates some level of degeneracy
between the first two squark generations, $\Delta \tilde
m^2_{12}/\tilde m^2\sim1/r_3$.

Throughout this paper we assume unsuppressed CP phases and $x \approx1$.
The supersymmetric contribution to $\epsilon_K$ that is proportional
to $(\delta^d_{LL})_{12}(\delta^d_{RR})_{12}\sim m_d/(r_3^2 m_s)$ is too large, unless
$r_3\gsim440\;({\rm TeV}/\tilde m)$ (or, equivalently, in the language used by CMSSM
practitioners, $m_{1/2}/m_0\gsim7$). Assuming that this is indeed the
case, the model is viable, and provides
\beq\label{eq:avail}
{\cal I}m(\delta_{LR})\sim 1.5\times10^{-4}\ \frac{\tilde a}{\tilde m}\
\frac{1\ {\rm TeV}}{\tilde m}.
\eeq
Comparing Eq. (\ref{eq:avail}) to Eq. (\ref{eq:desired}), we learn
that for supersymmetry to account for $\Delta A_{CP}$, the ratio
$(\tilde a/\tilde m)$ should be large. Taking $a_0,m_0$ and $m_{1/2}$
to stand for the Planck scale values of the $A$-terms, squark masses
and gluino mass, respectively, and using the approximations of Ref.
\cite{Hiller:2008sv}, we obtain
\beq
\frac{\tilde a}{\tilde m}\sim\frac{3a_0}{m_0\sqrt{1+8(m_{1/2}/m_0)^2}}
\to\left\{\begin{array}{cc}
 a_0/m_{1/2} & (m_{1/2}\gg m_0),\\
3a_0/m_{0}& (m_{1/2}\ll m_0).
\end{array}\right.
\eeq
Given that the $U(1)$ models are only viable if $m_{1/2}\gsim7m_0$,
the optimal enhancement occurs for $a_0>m_{1/2} \gg m_0$ which might,
however, lead to negative squark masses-squared.

The single $U(1)$ models lead to the following simple parametric
relation between the up and down sectors:
\beq \label{eq:uoverd}
\frac{(\delta^u_{LR})_{12}}{(\delta^d_{LR})_{12}}\sim\frac{m_c}{m_s}.
\eeq
The $(\delta^d_{LR})_{12}$ parameter is constrained, however, by
$\epsilon^\prime/\epsilon$: $(\delta^d_{LR})_{12}\lsim4\times10^{-5}\; ( \tilde m/{\rm
  TeV})$ (see \cite{Hiller:2008sv,Isidori:2010kg} and references
therein). We thus obtain an upper bound that is independent of the
flavor-diagonal scales,
\beq\label{eq:epsk}
(\delta^u_{LR})_{12}\lsim5\times10^{-4}\; \frac{\tilde m}{{\rm TeV}}.
\eeq

Moreover, the approximate symmetry relates $(\delta^u_{LR})_{12}$ to
flavor diagonal parameters,
\beq \frac{{\cal I}m(\delta^u_{LR})_{12}}{{\cal
    I}m(\delta^q_{LR})_{11}}\sim\frac{m_c|V_{us}|}{m_q},\ \ \ (q=u,d).
\eeq
Assuming phases of order one (which we must do to explain $\Delta
A_{CP}$), these flavor diagonal parameters are bounded by electric
dipole moment (EDM) constraints (see
\cite{Hiller:2008sv,Isidori:2010kg} and references therein),
$(\delta^u_{LR})_{11}\lsim3\times10^{-6}\;(\tilde m/{\rm TeV})$ and
$(\delta^d_{LR})_{11}\lsim2\times10^{-6}\;(\tilde m/{\rm TeV})$.  The
resulting bounds are
\beqa\label{eq:edmu}
(\delta^u_{LR})_{12}\lsim3\times10^{-4}\; \frac{\tilde m}{{\rm TeV}} ~~~  ({\rm from}\ (\delta^u_{LR})_{11}),\\
\label{eq:edmd}
(\delta^u_{LR})_{12}\lsim8\times10^{-5}\; \frac{\tilde m}{{\rm TeV}}~~~  ({\rm from}\ (\delta^d_{LR})_{11}).
\eeqa
We conclude that FN models with a single $U(1)$ are unlikely to
account for $\Delta A_{CP}\gg0.001$. Of course, since the FN mechanism
only dictates the parametric suppression, it is impossible to exclude
an accidental enhancement of $(\delta^u_{LR})_{12}$ by the order-one
coefficient.

Models with an FN symmetry $[U(1)]^2$ allow one to take advantage of
the holomorphicity of the superpotential to obtain vanishing entries
in the Yukawa and $A$ matrices and to strongly suppress entries in the
squark mass-squared matrices (compared to the single $U(1)$ case).
This feature was first employed in Refs.
\cite{Nir:1993mx,Leurer:1993gy} to align in a very precise way the
squark and quark mass matrices in the down sector.

The flavor structure of the Yukawa and $A$ terms in these models can be written
as follows:
\beqa
Y^d&\sim& \frac{\tilde M^{2d}_{LR}}{\tilde a v_d}\sim\left(\begin{array}{ccc}
y_d & 0 & y_b|V_{ub}| \\ 0 & y_s & y_b|V_{cb}| \\ 0 & 0 & y_b
\end{array}\right),\\
Y^u&\sim& \frac{\tilde M^{2u}_{LR}}{\tilde a v_u}\sim\left(\begin{array}{ccc}
y_u & y_c|V_{us}| & y_t|V_{ub}| \\ Y^u_{21} & y_c & y_t|V_{cb}| \\ Y^u_{31} & y_c/|V_{cb}| & y_t
\end{array}\right).
\eeqa
The four holomorphic zeros in the down sector are essential to obtain
an effective alignment \cite{Nir:2002ah}. The $(21)$ and $(31)$
entries in the up sector either (i) get their naive parametric
suppression (of order $y_u/|V_{us}|$ and $y_u/|V_{ub}|$,
respectively) or (ii) vanish.

In both cases, the contribution to $\epsilon_K$ from $(\delta^d_{LL})_{12}(\delta^d_{RR})_{12}$ for unsuppressed phases is too large, unless RGE generates degeneracy, $r_3\gsim 18\;({\rm TeV}/\tilde m)$.
(For moderate suppression of phases, $D^0-\overline{D}^0$ mixing provides the strongest constraint, requiring a milder degeneracy \cite{Gedalia:2012pi}:
In case (i),
the estimates (\ref{eq:LL}) and (\ref{eq:RR}) hold, and the contribution from $(\delta^u_{LL})_{12}(\delta^u_{RR})_{12}$ is too large, unless RGE generates degeneracy, $r_3\gsim 7\;({\rm TeV}/\tilde m)$ \cite{Hiller:2008sv}. In case (ii),
$(\delta^u_{RR})_{12}$ is suppressed compared to Eq. (\ref{eq:RR}),
and the contribution from $[(\delta^u_{LL})_{12}]^2$ only requires a
very mild degeneracy.)

In either case, the parametric suppression of
$(\delta^u_{LR})_{12}$ is as in Eq. (\ref{eq:LR}), and the numerical
estimate is as in Eq. (\ref{eq:avail}).  The parametric relation between the up and down sectors of Eq.~\eqref{eq:uoverd} does not hold in the $[U(1)]^2$ models since $(\delta^d_{LR})_{12}$ is further suppressed, and so the constraint of Eq.~\eqref{eq:epsk} does not hold. The constraints of Eqs.~(\ref{eq:edmu}) and (\ref{eq:edmd}) hold, leading to
the same conclusion as in the single $U(1)$ case: The parametric
suppression is such that the contribution to $\Delta A_{CP}$ falls an order of magnitude short compared to the benchmark value of Eq.~\eqref{eq:desired}. However, one cannot exclude the possibility that: (i) the large hadronic uncertainties in the EDM calculation are such that the bounds are weaker by an order of magnitude; or (ii) the order one uncertainty from the unknown coefficients provides further enhancement; or a combination of the above.

\subsection{Non-Abelian Symmetries}
In $U(2)$ models, the first two generations are in a doublet and the
third generation in a singlet of the $U(2)$ symmetry
\cite{Dine:1993np,Pomarol:1995xc,Barbieri:1995uv,Barbieri:1997tu}.
With a two-stage symmetry breaking, the structure of the Yukawa, $A$
and $\tilde M^2$ matrices is as follows (see, for example,
\cite{Barbieri:1997tu}):
\beqa
Y^q&\sim&\frac{\tilde M^{2q}_{LR}}{\tilde a v_q}\sim
\left(\begin{array}{ccc}
0 & \epsilon_1 & 0 \\ -\epsilon_1 & \epsilon_2 & \epsilon_2 \\
0 & \epsilon_2 & 1\end{array}\right),\no\\
\tilde M^{2q}_{NN}&\sim&\left(\begin{array}{ccc}
m_1^2 & 0 & 0 \\ 0 & m_1^2(1+\epsilon_2^2) & \epsilon_2 m_4^{2*} \\
0 & \epsilon_2 m_4^2 & m_3^2\end{array}\right).
\eeqa
As in the Abelian case, all non-vanishing entries have unknown
coefficients of order one, but $Y^q_{12}=-Y^q_{21}$ and there are
relations for the $\tilde M^2_{NN}$ matrices that follow from
hermiticity.

As concerns the $\delta_{LR}^q$ parameters, their parametric
suppression is similar to the $U(1)$ model. Hence, Eqs. (\ref{eq:LR}),
(\ref{eq:epsk}), (\ref{eq:edmu}) and (\ref{eq:edmd}) all hold.

The main phenomenological difference of the $U(2)$ model with respect
to the $U(1)$ model is that the first two squark generations are
quasi-degenerate already at the mediation scale, with a mass splitting
$\Delta\tilde m^2_{12}/\tilde m^2\sim\epsilon_2^2\sim10^{-3}$. Hence,
the model is viable even without invoking flavor-universal RGE
effects.

A flavor $[U(2)]^3$ symmetry \cite{Barbieri:2011ci} is motivated by
the tension between the measured value of the CP asymmetry $S_{\psi
  K}$, and its theoretical value in the Standard Model derived from a
global CKM fit. To alleviate this tension, a new physics contribution
to $B^0-\overline{B}^0$ mixing of order 10 percent of the total
amplitude is required. In a $U(2)$ model, such a contribution entails
a contribution to $\epsilon_K$ of order 100 percent, which is
unacceptable.  A $U(2)_Q\times U(2)_U\times U(2)_D$ model, with
minimal spurion content -- $V(2,1,1)$, $\Delta Y_u(2,\bar2,1)$ and
$\Delta Y_d(2,1,\bar2)$ -- allows one to suppress the contribution to
$\epsilon_K$.

The structure of the Yukawa and $A$ matrices is as follows
\cite{Barbieri:2011ci}:
\beqa
Y^q&\sim&\frac{\tilde M^{2q}_{LR}}{\tilde a v_q}\sim
y_{q_3}  \left(\begin{array}{cc}
\Delta Y_q & x_q V \\ 0 & 1\end{array}\right),
\eeqa
where $y_{q_3}=y_t(y_b)$ for $q=u(d)$, $x_q$ is a complex free
parameter of order one, $\Delta Y_q$ is a $2\times2$ matrix, and $V$
is a $2\times1$ vector.

This structure is quite unique in that one and the same spurion,
$\Delta Y_q$, determines the structure of the $2\times2$ upper left
block of both $Y^q$ and $\tilde M^{2q}_{LR}$. Consequently, to leading
order in the breaking parameters, $(\delta^q_{LR})_{12}=0$, and the
supersymmetric contribution to $\Delta A_{CP}$ vanishes. Corrections
to $\delta_{LR}$ arise at the order $y_c y_t V_{cb}^* V_{ub}$ and are
negligible.

\subsection{Hybrid Mediation}
In models of hybrid mediation, the dominant source of supersymmetry
breaking is MFV, but there are non-negligible contributions from
Planck scale physics that do not obey the MFV principle. Examples
include high-scale gauge mediation \cite{Feng:2007ke} and a class of
models with anomaly mediation \cite{Gross:2011gj}.  At the messenger
scale, the relative size between the soft masses-squared arising from
gravity and MFV physics is given by $r\sim \tilde m_{\rm
  grav}^2/\tilde m_{\rm MFV}^2$.

The $c \to u$ couplings are given by
\beqa\label{eq:RRh}
(\delta^u_{LL})_{12}&\sim&\frac{r |V_{us}|}{r_3},\label{eq:LLh}\\
(\delta^u_{RR})_{12}&\sim&\frac{r\, m_u}{r_3 m_c|V_{us}|},
\eeqa
and the expressions for $\delta_{LR}$ remain as in the pure gravity case, Eqs. (\ref{eq:LR}) and (\ref{eq:avail}).

The EDM constraints of Eqs.~(\ref{eq:edmu}) and (\ref{eq:edmd}) hold.
One can now ask what further constraints arise when linking the
trilinear terms and the soft masses as is characteristic in hybrid
models. If all gravity soft terms are dictated by a single scale, then
$a_0\sim \sqrt r\; \tilde m_{\rm MFV}$, where $\tilde m_{\rm MFV}$ is
the typical messenger scale MFV soft mass. The relevant combination
entering $\delta_{LR}$ is then $\tilde a/\tilde m \lesssim 3
\sqrt{r/r_3}$, where the numerical factor stems from RGE and is
largest for high scale mediation. In the following we analyze this
single gravity scale scenario in the FN context.

In models with a single $U(1)$ and order one phases kaon mixing requires $r/r_3 \lesssim
0.002\;(\tilde m/{\rm TeV})$~\cite{Hiller:2008sv}.  Therefore,
\beq\label{eq:availh}
{\cal I}m(\delta_{LR})\lesssim 0.2 \times10^{-4}\
\sqrt{\frac{{\rm TeV}}{\tilde m}},
\eeq
a stronger constraint than the EDM bound of Eq.~(\ref{eq:edmd}).

In $[U(1)]^2$ models the kaon system constrains $r/r_3 \lesssim
0.06\;(\tilde m/{\rm TeV})$, and so
\beq\label{eq:availhu12}
{\cal I}m(\delta_{LR})\lesssim 1 \times10^{-4}\
\sqrt{\frac{{\rm TeV}}{\tilde m}},
\eeq
close to the upper bound from EDMs, Eq.~(\ref{eq:edmd}). We note that it is possible in specific $U(1)^2$ models to further suppress the contribution to the kaon system, such that the strongest bound comes from $D^0-\overline D^0$ mixing and gives $r/r_3\lsim0.8\;(\tilde m/{\rm TeV})$, relaxing the constraint~\eqref{eq:availhu12} by a factor of 4.

Both $U(2)$ and $[U(2)]^3$ models do not require further flavor
suppression, and are viable for $r \lesssim 1$, with predictions as in
the non-hybrid models.

We conclude that hybrid models with a $[U(1)]^2$, $U(2)$ or $U(2)^3$
symmetry generate $\Delta A_{CP}$ of the same size as non-hybrid
models. The size of $\delta_{LR}$ allowed by hybrid models with a
single $U(1)$ is somewhat smaller.

\section{$A$ terms and the lightest Higgs mass}
In supersymmetry $\Delta A_{CP}$ can be interpreted via the left-right
mixing $\delta_{LR}$ which requires unsuppressed trilinear couplings
with respect to the squark masses, $\tilde a/\tilde m\sim {\cal
  O}(1)$, see Eq.~\eqref{eq:avail}.  At the same time the recent hints
from ATLAS and CMS of a neutral Higgs boson with mass near
125~GeV~\cite{:2012si} implies that the stops -- if not decoupled --
are largely mixed as well~\cite{Heinemeyer:2011aa,Draper:2011aa}:
\beq \label{eq:stop-mixing}
|A_t/y_t -\mu/\tan \beta| \gtrsim M_S,
\eeq
where $M_S$ denotes the geometric mean of the stop masses.
In the FN models, where the flavor structure of the $A$ terms is
parametrically similar to that of the Yukawas, $A \sim Y$, the stop
$A$ terms at the weak scale can be written as~\cite{Martin:1993zk}
\beq
A_t/y_t \simeq \tilde a -\Delta a, ~~ \Delta a=y_t^2 b a_0+m_{1/2}c,
\eeq
with positive RGE-induced coefficients $b,c$ of order one.

For positive $\mu$ or sufficient $\tan \beta$ suppression one obtains
from Eq.~(\ref{eq:stop-mixing}) a lower bound that supports a sizeable
supersymmetric $\Delta A_{CP}$,
\beq \label{eq:atermlimit}
\tilde a/\tilde m  \gtrsim M_S/\tilde m
\eeq
for $A_t >0$ and unsplit spectrum where the stops are not too far away from the other squarks. Negative $A_t<0$ can arise in scenarios with tiny or
vanishing $a_0$ such as gauge mediation, which lead to acceptable
phenomenology only for sufficiently large gluino masses.

While the Higgs signal needs to be consolidated, it is interesting
that if confirmed, the current mass of $\sim 125$~GeV points to a
similar region in supersymmetric parameter space as the interpretation
of $\Delta A_{CP}$.

\section{Conclusions}
Supersymmetric models can contribute to direct CP violation in singly
Cabibbo suppressed $D$ decays at the level observed by experiments,
$\Delta A_{CP}^{\rm SUSY}\sim0.006$, without conflicting with
phenomenological constraints from $D^0-\overline{D}^0$ mixing or
$\epsilon^\prime/\epsilon$.  This is naturally the case if the flavor
changing parameter $(\delta^u_{LR})_{12}$, generated by trilinear
scalar couplings, is of order $10^{-3}$.

In minimally flavor violating supersymmetric models, such as those of
gauge mediation and anomaly mediation, $(\delta^u_{LR})_{12}$ is
orders of magnitude too small. Thus, to account for $\Delta A_{CP}$,
one has to go beyond minimal flavor violation. We examined models
where the flavor structure of the soft breaking terms is dictated by
an approximate flavor symmetry.

We found that quite generically in such models, $(\delta^u_{LR})_{12}$
is flavor-suppressed by $(m_c|V_{us}|/\tilde m)$, which is of order
a few times $10^{-4}$. There is however additional dependence on
the ratio between flavor-diagonal parameters, $\tilde a/\tilde m$,
and on unknown coefficients of order one, that can provide enhancement
by a factor of a few.

In most such models, however, the selection rules that set the flavor
structure of the soft breaking terms, relate $(\delta^u_{LR})_{12}$ to
$(\delta^d_{LR})_{12}$ and to $(\delta^{u,d}_{LR})_{11}$, which are
bounded from above by, respectively, $\epsilon'/\epsilon$ and EDM constraints.
Since both $\epsilon'/\epsilon$ and EDMs suffer from hadronic uncertainties, small
enhancement due to the flavor-diagonal supersymmetric
parameters cannot be ruled out. Additionally, it is still possible that $(\delta^u_{LR})_{12}$ is
accidentally enhanced by the order one coefficient.

Chirality-flipping couplings between the first and second generation up squarks can effectively arise also via chirality-flipping in the third generation~\cite{Giudice:2012qq}. In all flavor models considered here, the effective $\delta_{LR}$ generated is at most parametrically of the same order as the direct contribution, accompanied by an additional $1/r_3^2$. For $U(1)$ and $[U(1)]^2$ models this provides extra suppression, while for $U(2)$ models the effective contribution to $\delta_{LR}$ is flavor-suppressed with respect to the direct one. In any event, the constraints remain the same and the analysis stands.

We conclude that it is possible to accommodate $\Delta
A_{CP}\sim0.006$ in supersymmetric models that are non-minimally
flavor violating, but -- barring hadronic enhancements in charm decays
-- it takes a fortuitous accident to lift the supersymmetric
contribution above the permil level.

\section{Acknowledgments}
We thank Gino Isidori for useful discussions.
This project is supported by the German-Israeli foundation for
scientific research and development (GIF). YN is the Amos de-Shalit
chair of theoretical physics and supported by the Israel Science Foundation.


\end{document}